\documentclass[]{emulateapj}

\newcommand{\FeI}{\ion{Fe}{1}}

\newcommand{\CaII}{\ion{Ca}{2}}
\usepackage{mathptmx}
\usepackage{courier}
\usepackage{graphicx}
\usepackage{colordvi}
\usepackage{natbib}

\newcommand{\figdir}[1]{./#1}

\shortauthors{Bellot Rubio, Tritschler, Mart\'{\i}nez Pillet}
\shorttitle{Spectropolarimetry of a decaying sunspot penumbra}

\begin{document}

%
%

\title{Spectropolarimetry of a decaying sunspot penumbra}
\author{L.R.~Bellot Rubio}
\affil{Instituto de Astrof\'{\i}sica de Andaluc\'{\i}a (CSIC), Apartado 3004,
18080 Granada, Spain; lbellot@iaa.es}
\and
\author{A.~Tritschler}
\affil{National Solar Observatory/Sacramento Peak\footnote{Operated by the %
       Association of Universities for Research in Astronomy, Inc. (AURA), %
       for the National Science Foundation},
       P.O.~Box 62, Sunspot, NM 88349, U.S.A.; ali@nso.edu}
\and
\author{V.\ Mart\'{\i}nez Pillet}
\affil{Instituto de Astrof\'{\i}sica de Canarias, C/Via L\'actea s/n, 38200 La
Laguna, Tenerife, Spain; vmp@iac.es}

%
%

\begin{abstract}
We report on high angular resolution, high precision spectropolarimetric
measurements of a decaying sunspot. The spot gradually lost its penumbra
during the course of three days. In the late stages of evolution where the
only remnant of the spot is a naked umbra, we find small-scale inhomogeneities
in the magnetic canopy surrounding it. The inhomogeneities are observed as
finger-like structures of weak and nearly horizontal magnetic fields extending
1-2\arcsec\/ from the outer border of the umbra. These fields are not
associated with filamentary structures in continuum intensity, nor with
conspicuous Evershed flows. The Stokes profiles emerging from the fingers
exhibit blueshifts which we interpret as upward motions. This previously
unknown fine structure may be related to penumbral field lines that no longer
carry strong Evershed flows and rise to the chromosphere, producing the
disappearance of the penumbra at photospheric levels.
\end{abstract}


%
%

\section{Introduction}\label{sec:intro} 

The formation of sunspot penumbrae is a relatively well known process,
both from an observational and theoretical point of view. Pores
develop penumbrae suddenly, with no obvious gradual increase of the
magnetic field inclination and little (if any) delay between the
appearance of Evershed flows and inclined fields
\citep{leka+skumanich1998}. From the theoretical side, it has been
suggested that penumbra formation occurs when the magnetic field at
the outer edges of pores reaches inclinations of about $45^\circ$
\citep{rucklidge+etal1995, tildesley+weiss2004}.

The disappearance of the penumbra, in contrast, still needs to be
understood.  Observations indicate that penumbral decay is a slow
(several day) process, but to our knowledge no spectropolarimetric
measurements of it have been presented so far. Studies of decaying
active regions often concentrate on other aspects such as moving
magnetic features \citep[e.g.,][]{zhangetal92} or the moat flow
\citep[e.g.,][]{dengetal08}. The disappearance of sunspot penumbrae is
an important phenomenon which may partly explain the removal of flux
from mature spots \citep[cf.][and references therein]{vmp2002}.  If
so, it would play a crucial role in the activity cycle of the Sun.
Penumbral decay may also be a source of localized chromospheric and
coronal heating.  Indeed, energetic flares in active regions have been
related to the disappearance of sunspot penumbrae
\citep{wang+etal2004, deng+etal2005, alberto+vmp2007}.

In this paper we analyze high angular resolution spectropolarimetric
measurements of a decaying sunspot. We concentrate on the phase in
which the sunspot lost all visible evidence of its penumbra. A Stokes
inversion technique is used to determine magnetic field inclinations,
field strengths, and line-of-sight (LOS) velocities. We find evidence
for small-scale inhomogeneities in the magnetic canopy of the spot
when it consists only of a naked umbra. The inhomogeneities are
observed as horizontal field lines cospatial with blueshifted Doppler
signals. The blueshifts could indicate the occasional establishment of
an inward oriented Evershed flow.  Alternatively, these structures may
be related to penumbral field lines that no longer carry an Evershed
flow. In the absence of strong material flows, the field lines could
rise to the chromosphere by buoyancy, which would be observed in the
photosphere as the disappearance of the penumbra.

The paper is organized as follows: in \S~\ref{sec:obs} and
\ref{sec:analysis} we describe the observations and data analysis.
The results are presented in \S~\ref{sec:results}.  In
\S~\ref{sec:discussion}, we speculate on scenarios that might explain
the observations. Finally, \S~\ref{sec:conclusions} summarizes our
main conclusions.

%
%

\section{Observations}\label{sec:obs}

The spectropolarimetric observations analyzed here were obtained with
the Diffraction Limited Spectro-Polarimeter \citep[DLSP;
][]{sankar+etal2004} at the NSO Dunn Solar Telescope (DST) of
Sacramento Peak, New Mexico.  The DLSP was operated in high-resolution
mode together with the high-order adaptive optics system of the DST
\citep{rimmele2004}. The slit width of 12\,microns matched the
detector image scale along the spectrograph slit (0.090\arcsec \,
pixel$^{-1}$). The solar spectrum was sampled from 630.0\,nm to
630.4\,nm with 2.1\,pm\,pixel$^{-1}$. This wavelength range contains
three \ion{Fe}{1} and one \ion{Ti}{1} magnetically sensitive lines
(cf.\ Table~\ref{tab:lines}).

\begin{table}
\caption{Atomic parameters of the lines observed with the DLSP 
\label{tab:lines}}
\tabcolsep 1.5em
\begin{tabular}{l c c c c}
\tableline
Line & $\lambda_{0}$ & $\chi$ & $\log{ gf}$ & Transition \\
 & [nm] & [eV]  & &  \\
\tableline
\ion{Fe}{1}   & $630.1501^a$    & $3.65$ & $-0.72$ & $^{5}P_{2}$--$^{5}D_{2}$ \\
\ion{Fe}{1}   & $630.2494^a$    & $3.69$ & $-1.24$ & $^{5}P_{1}$--$^{5}D_{0}$ \\
\ion{Fe}{1}   & $630.3460^b$  & 4.32       & $-2.55$ & $^{5}G_{6}$--$^{5}G_{5}$  \\
\ion{Ti}{1}   & $630.3753^b$  & 1.45       & $-1.44$ & $^{3}F_{3}$--$^{3}G_{3}$ \\  
\tableline
\end{tabular}
\tablerefs{{\em a:} \cite{1994ApJS...94..221N}; {\em b:} \cite{borrero+bellotrubio2002}}
\vspace*{1.5em}
\end{table}

\begin{figure*}
\begin{center}
  \scalebox{0.5}{\includegraphics[bb=1 1 226 226,clip]{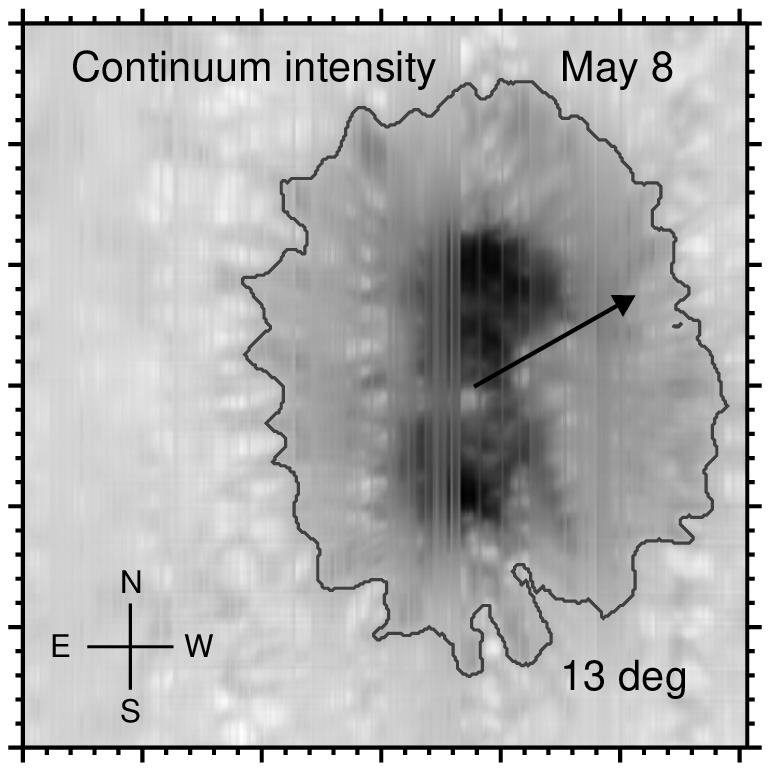}}
  \scalebox{0.5}{\includegraphics[bb=1 1 226 226,clip]{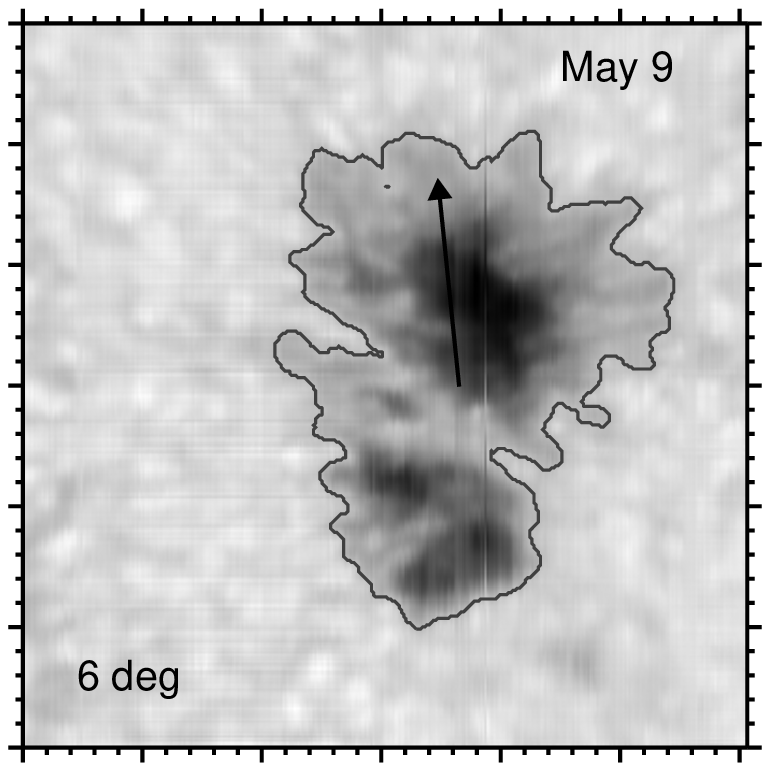}}
  \scalebox{0.5}{\includegraphics[bb=1 1 226 226,clip]{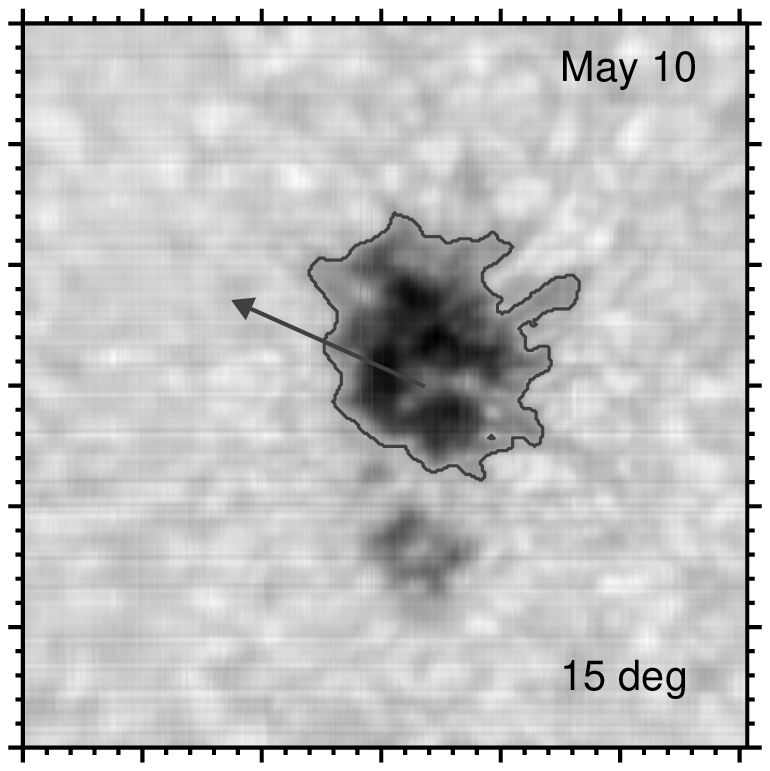}}
  \scalebox{0.5}{\includegraphics[bb=1 1 226 226,clip]{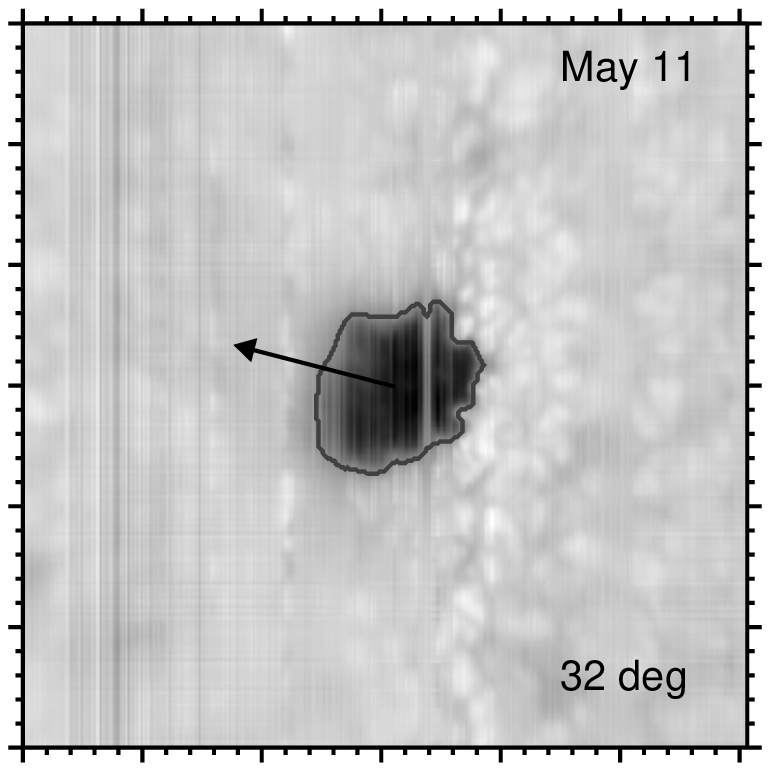}}\\
  \scalebox{0.5}{\includegraphics[bb=1 1 226 226,clip]{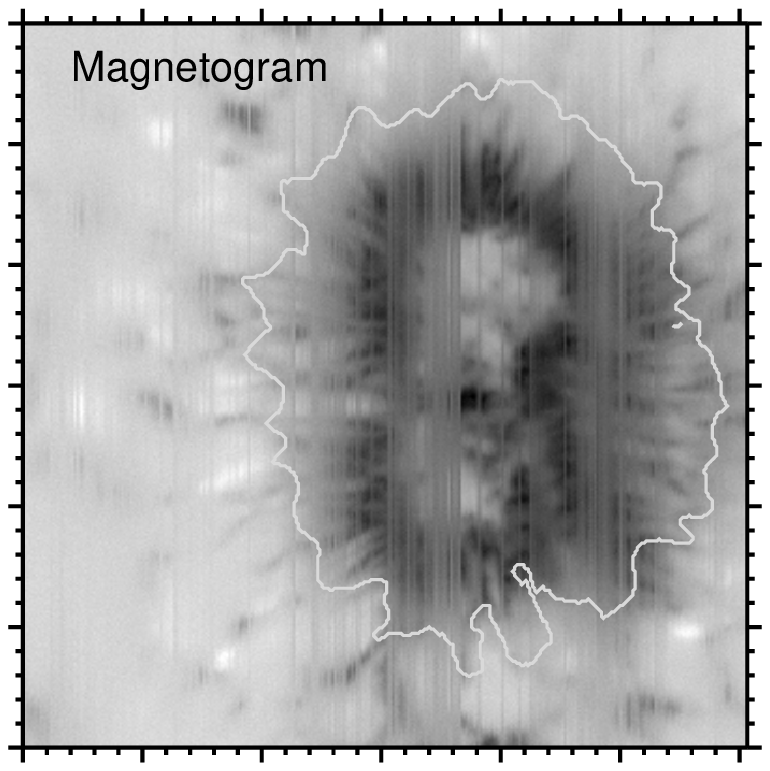}}
  \scalebox{0.5}{\includegraphics[bb=1 1 226 226,clip]{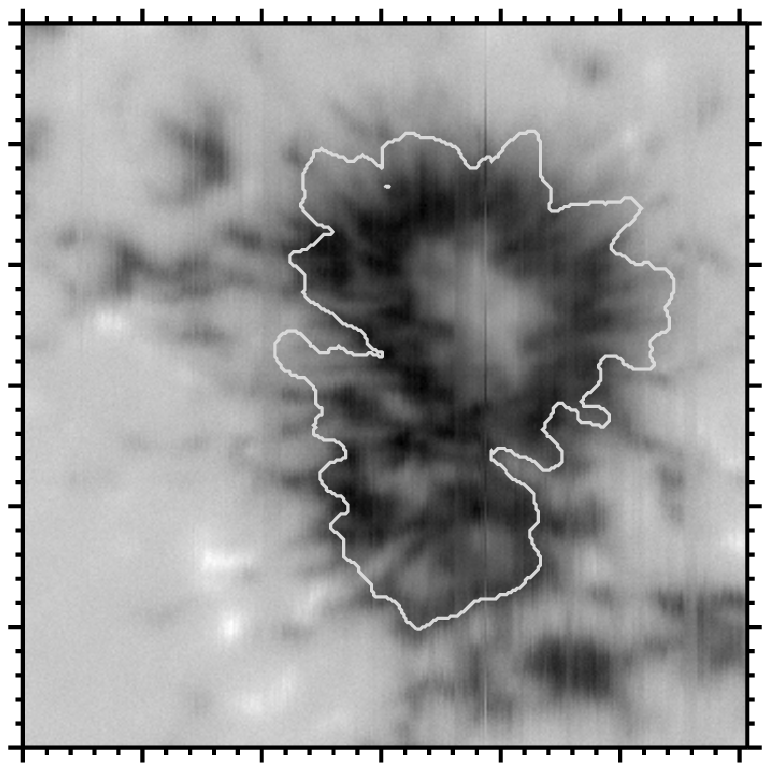}}
  \scalebox{0.5}{\includegraphics[bb=1 1 226 226,clip]{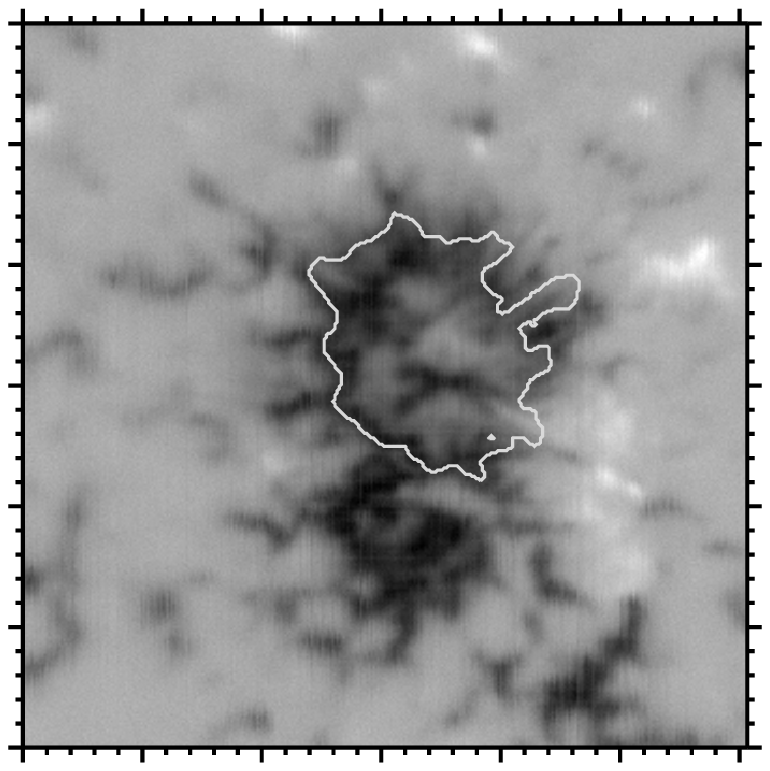}}
  \scalebox{0.5}{\includegraphics[bb=1 1 226 226,clip]{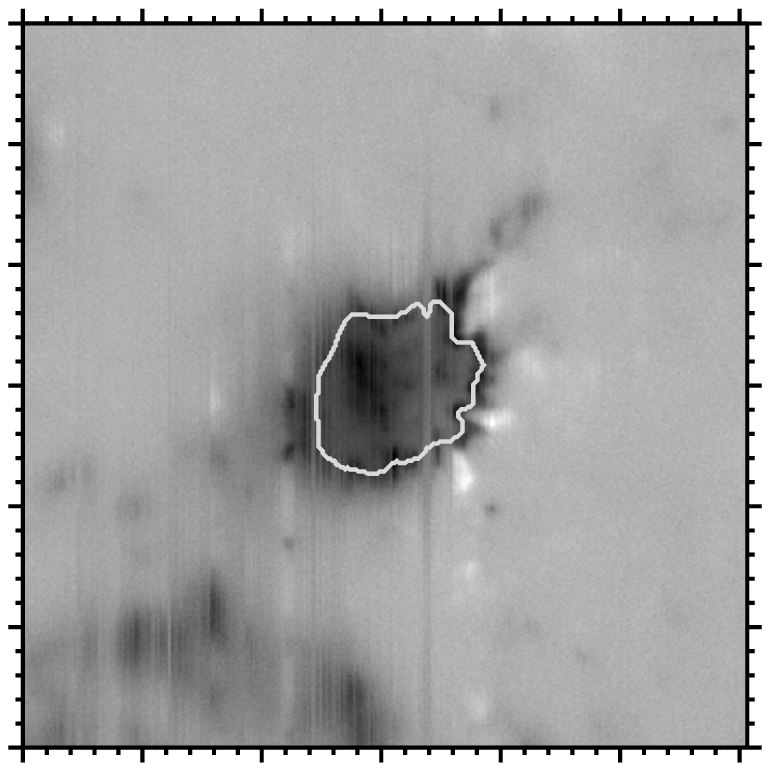}} \\
  \scalebox{0.5}{\includegraphics[bb=1 1 226 226,clip]{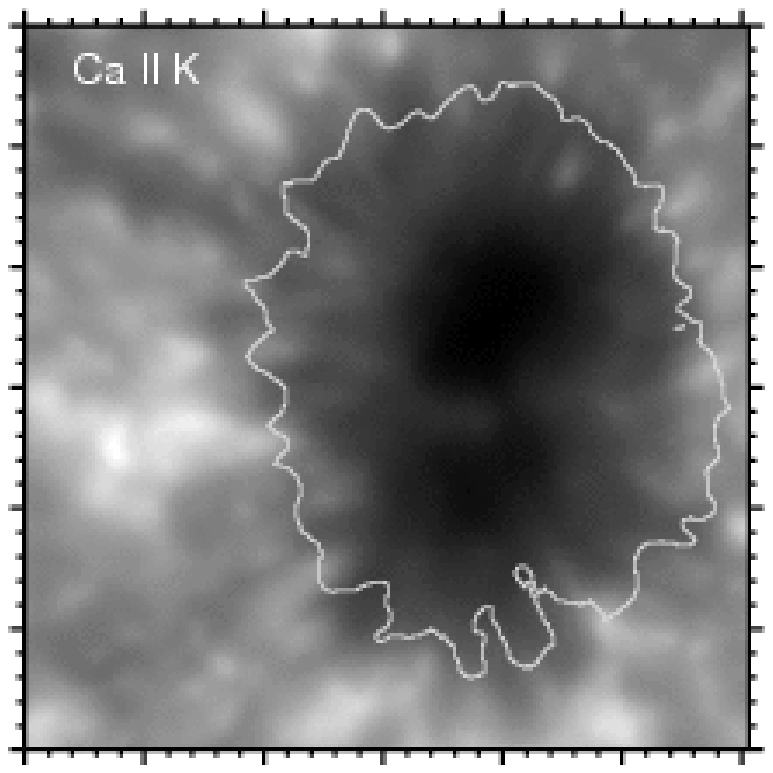}}
  \scalebox{0.5}{\includegraphics[bb=1 1 226 226,clip]{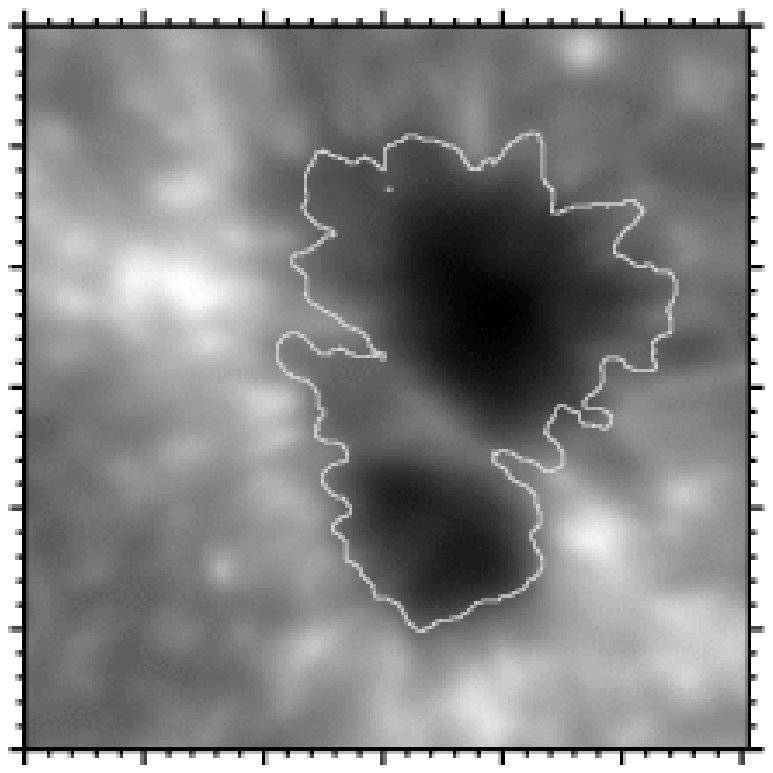}}
  \scalebox{0.5}{\includegraphics[bb=1 1 226 226,clip]{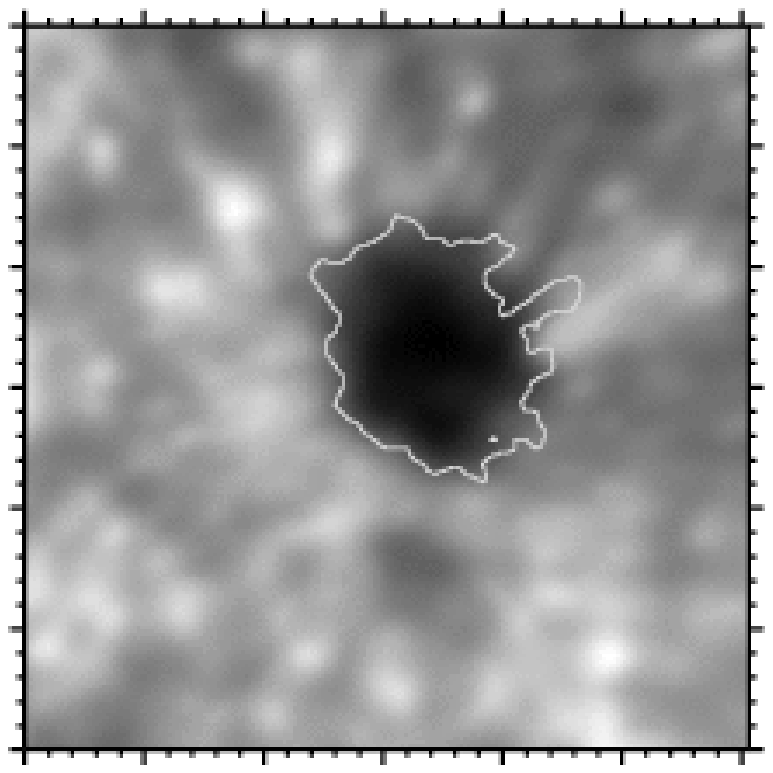}}
  \scalebox{0.5}{\includegraphics[bb=1 1 226 226,clip]{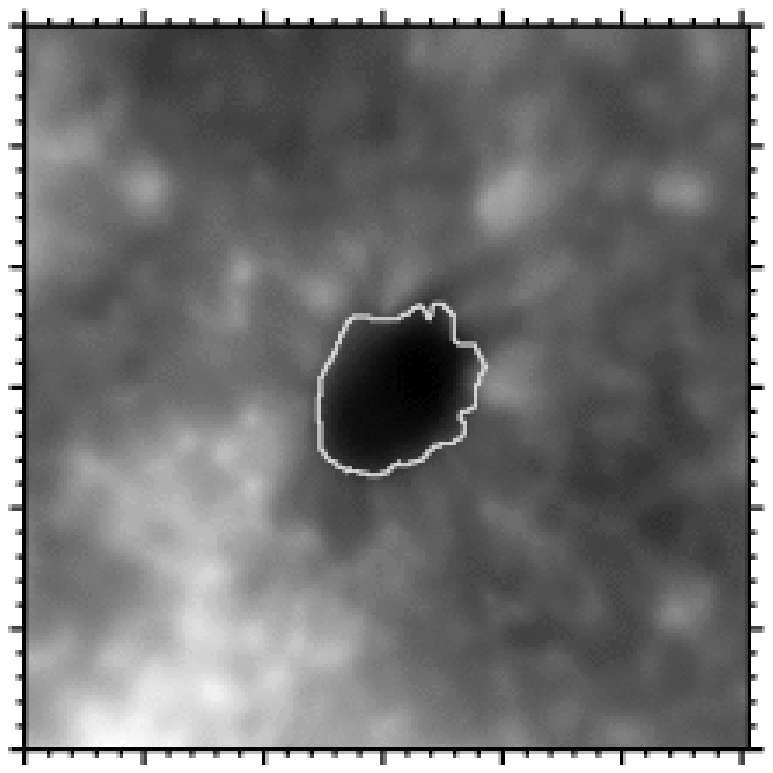}} \\  
  \caption{Evolution of the spot during four consecutive days. {\em Top:}
  Continuum intensity at 630.2 nm. {\em Middle:} Stokes $V$ signal observed in
  the blue wing of \ion{Fe}{1} 630.25 at $-9$~pm from line center. {\em
  Bottom:} \CaII~K filtergrams. Tick marks are separated by 1\arcsec. The
  heliocentric angle is indicated in the continuum intensity images. The arrows
  mark the direction to disk center. Contours represent the visible border of
  the spot. }
  \label{fig:obs}
  \epsscale{1.0}
\end{center}
\end{figure*}

By moving the slit across the solar surface with steps of 0\farcs0893
for a total of 337 positions, two dimensional maps of AR NOAA 10880
were generated on four consecutive days, from May 8 to May 11,
2006. The maps cover a field of view of $30.1\arcsec \times
57.2\arcsec$. The slit was oriented along solar North-South and the
scan direction was solar West-East. At every slit position, 16 images
of 30\,ms exposure time were accumulated for each of the four
individual modulation states, leading to a total integration time of
2\,s and completion of a full map within 28 minutes. The seeing
conditions were quite variable on May 8 and May 11 and more stable on
May 9 and May 10, as can be seen in the continuum maps of
Fig.~\ref{fig:obs}.  Mostly the variable seeing and weather
conditions, but also the time consuming calibration measurements
needed to correct for instrumental polarization, were the prime
limitations during this campaign, leading to one map per day only.
Although the observations on May 8 and May 11 were most affected by
variable conditions, these scans also featured lasting moments of
excellent seeing during which an angular resolution of about 0\farcs4
was attained.

We have applied standard reduction procedures to the data, including dark
subtraction, flatfielding, and removal of instrumental
polarization. After correction, the noise level in the Stokes profiles
is $10^{-3} \, I_c$ as measured in the continuum of $Q$, $U$, and $V$.
We perform an absolute velocity calibration by taking advantage of the
telluric O$_2$ lines present in the observed wavelength range
\citep[see e.g. ][]{martinezpillet+lites+skumanich1997}.

The spectropolarimetric measurements were complemented with near
synchronous G-band and \CaII~K imaging. We used narrow-band
interference filters centered at 430.55\,nm (0.92\,nm FWHM) for the G
band and at 393.34\,nm (0.1\,nm FWHM) for \CaII~K, with exposure times
of 10\,ms and 100\,ms, respectively. Pixel sizes correspond to
0\farcs03$\times$0\farcs03 in the G-band channel and to
0\farcs12$\times$0\farcs12 in the \CaII~K channel. The G-band images
have been reconstructed using a speckle masking technique implemented
by \cite{woeger2007}. The spatial resolution of the reconstructed
filtergrams is estimated to be 0\farcs16, close to the diffraction
limit of the DST at this wavelength (0\farcs12). The \CaII~K images
reach a spatial resolution of some 0\farcs6. The alignment between
imaging data and spectropolarimetric observations was carried out with
respect to the DLSP continuum maps.  We estimate that residual
rotation between the data sets is less than a fraction of a degree
($<0.1^\circ$). Image scale differences after re-scaling of the 
G-band and the \CaII~K pixels are smaller than 1\%, and residual 
offsets between the images are less than a fraction of a 
pixel ($<0.1$ pixel).


%
%

%

\begin{figure*}[t]
  \epsscale{0.36}
  \plotone{\figdir{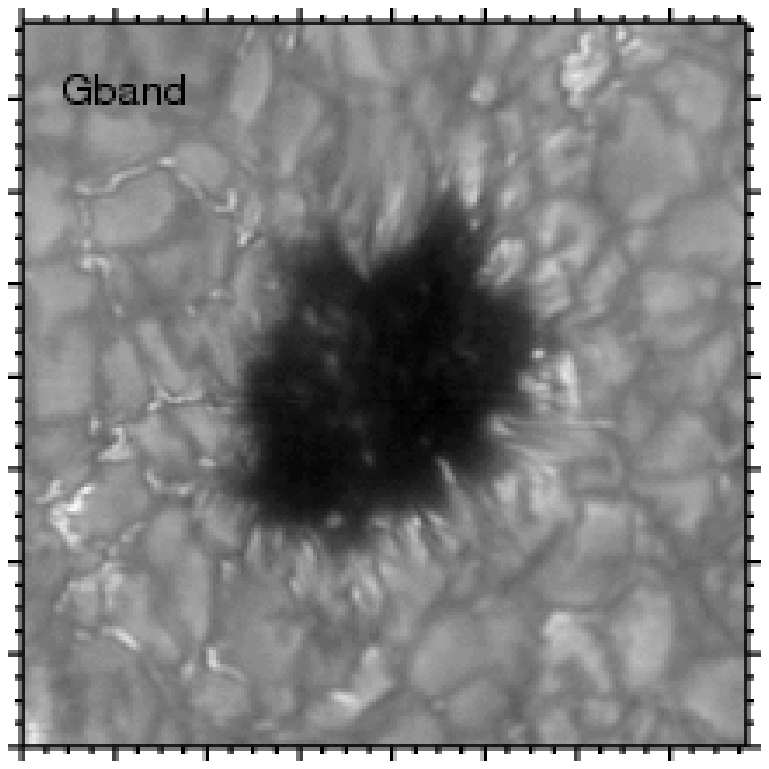}}
  \plotone{\figdir{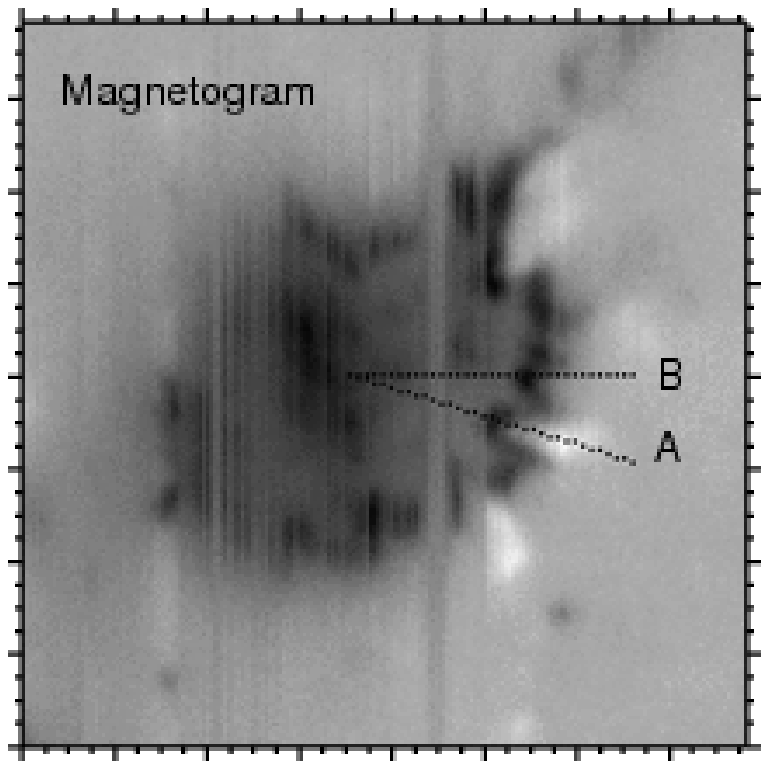}} \\
  \plotone{\figdir{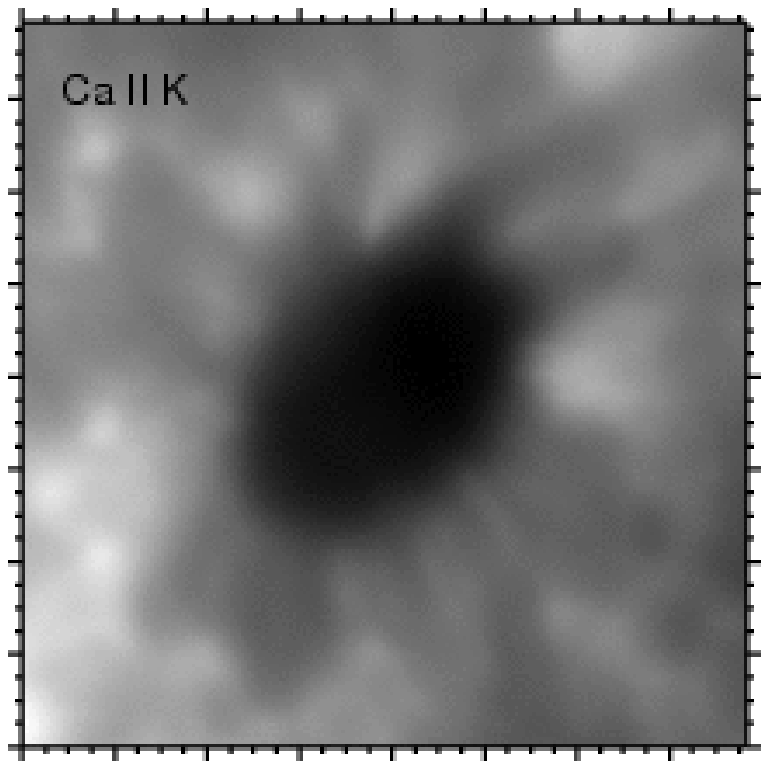}}
  \plotone{\figdir{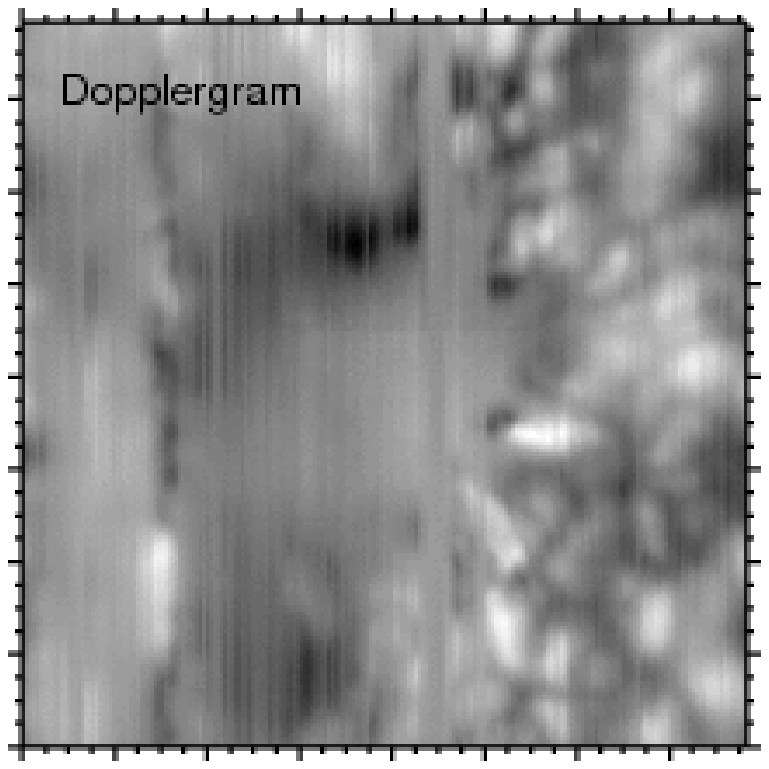}}
  \caption{Late evolutionary state of AR 10880 on May 11, 2006.  {\em
Clockwise, starting from upper left:} G-band filtergram, magnetogram,
Dopplergram, and \CaII~K filtergram.  Tickmarks are separated by
0\farcs5. Bright and dark areas in the Dopplergram move toward the observer
and away from it, respectively. The disappearing spot no longer has a
penumbra. Cuts A and B in the magnetogram sample one finger and the
undisturbed magnetic canopy, respectively.}
  \label{fig:physmaps}
  \epsscale{1.0}
\vspace*{1em}
\end{figure*}

\begin{figure*}
\vspace*{-6em}
  \epsscale{.236}
  \plotone{\figdir{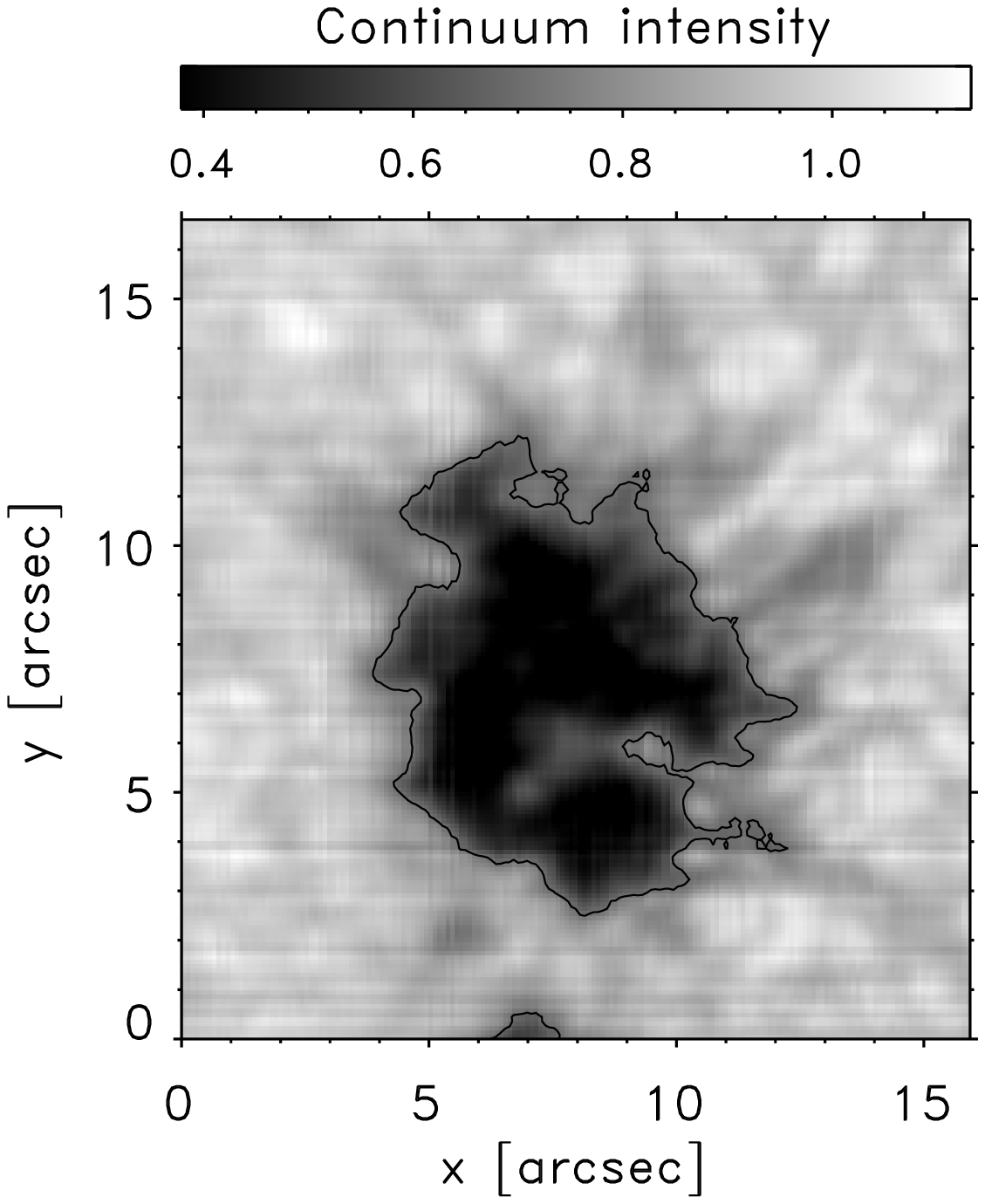}}
  \plotone{\figdir{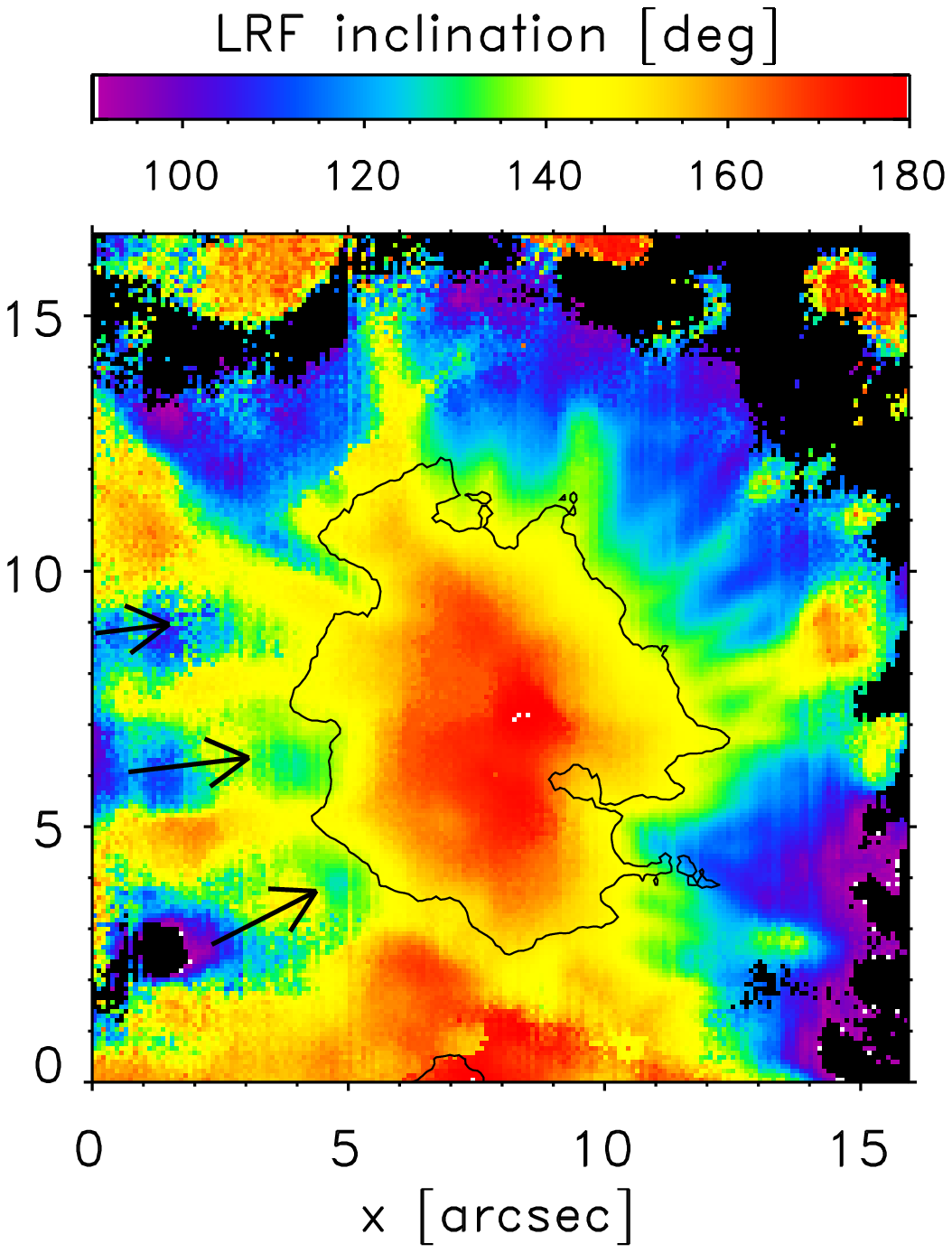}}
  \plotone{\figdir{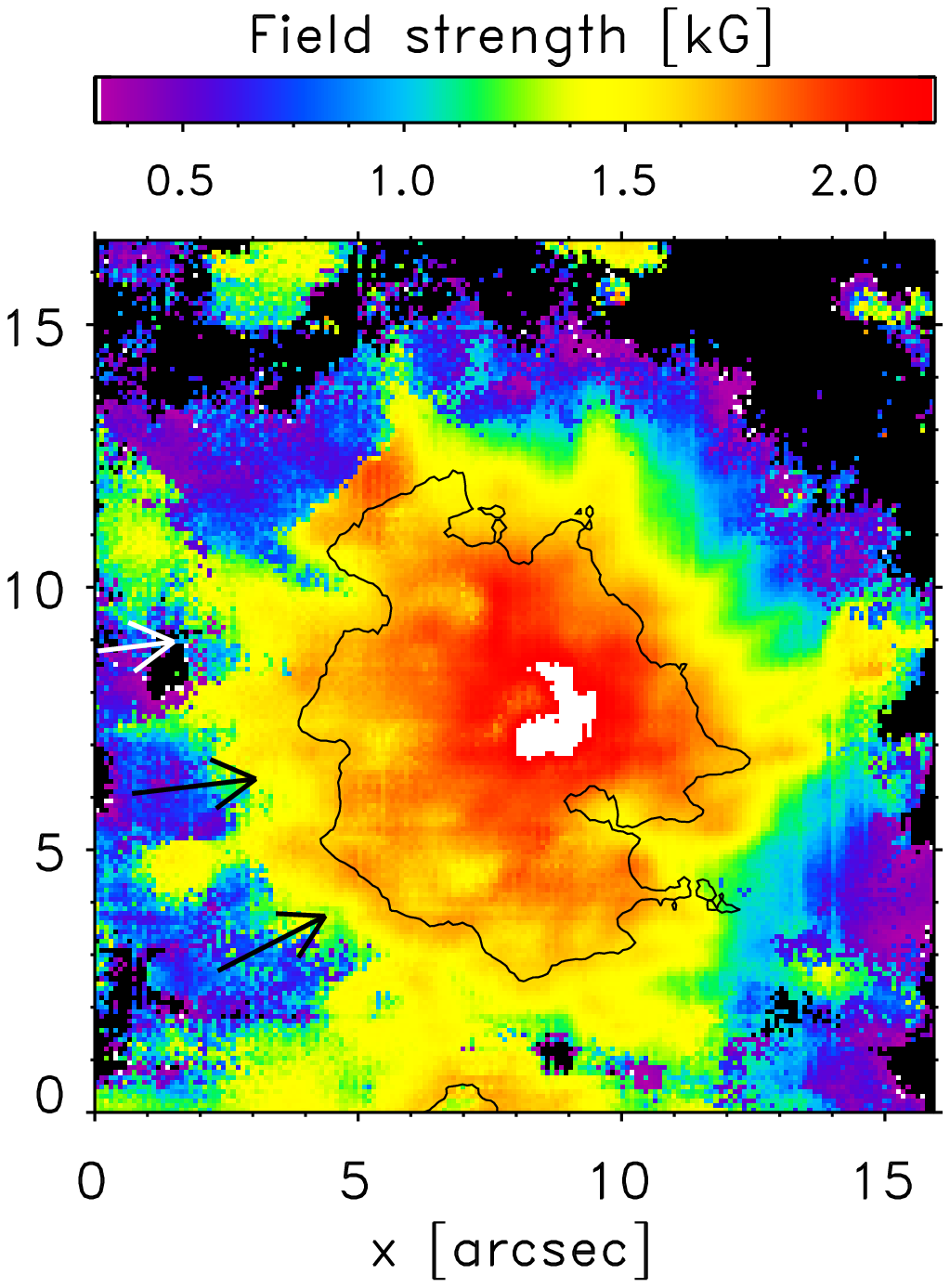}}
  \plotone{\figdir{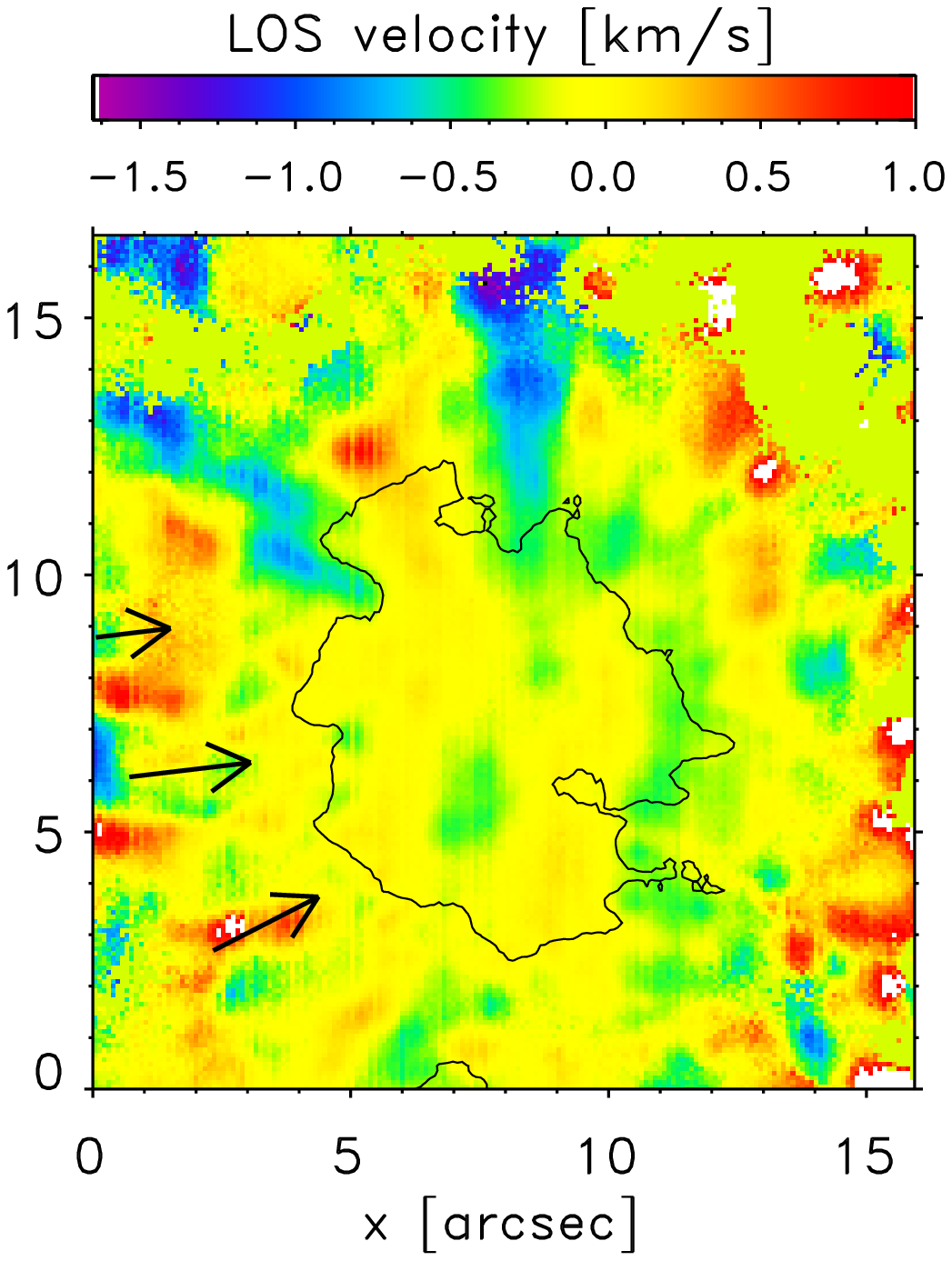}} \\
  \plotone{\figdir{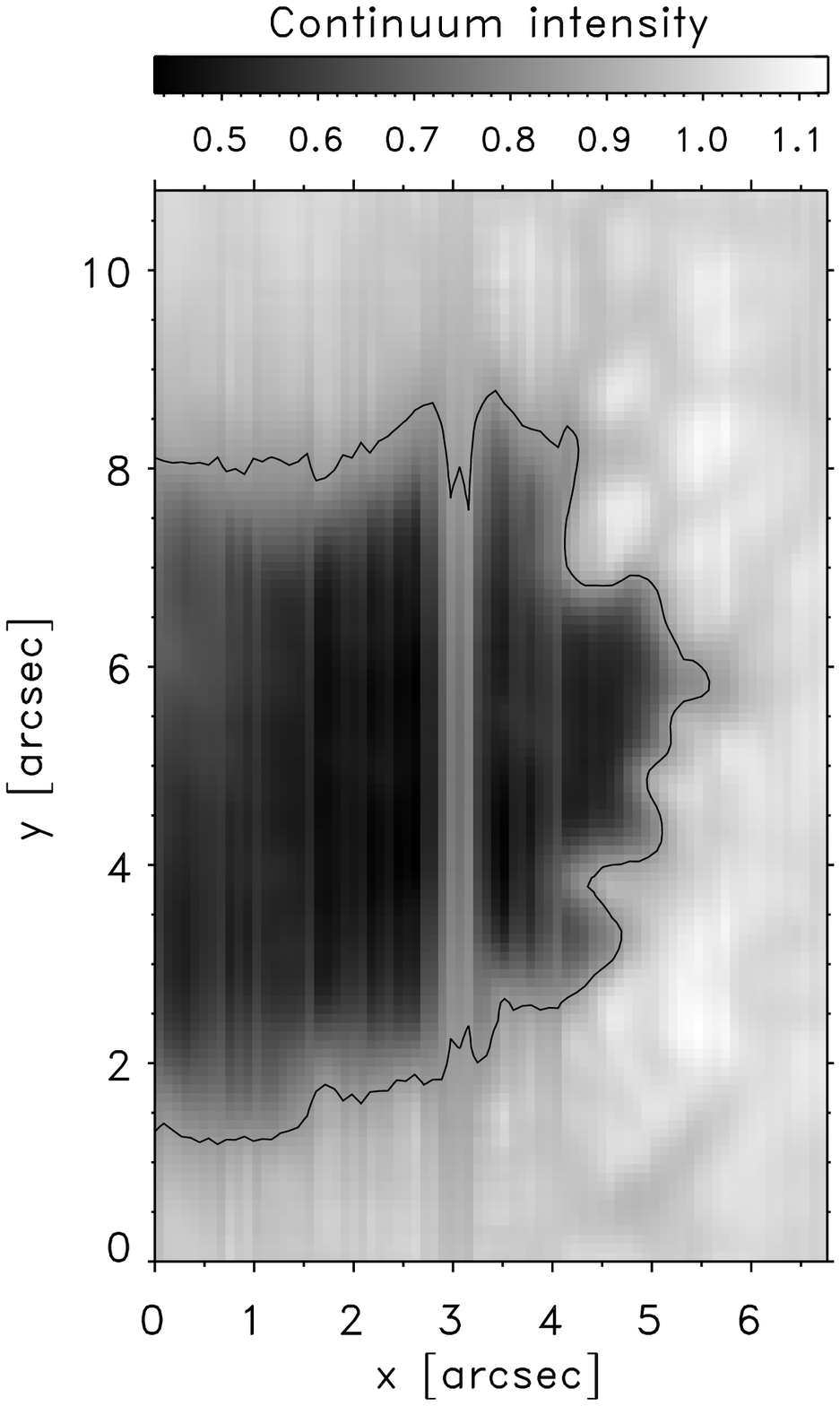}}
  \plotone{\figdir{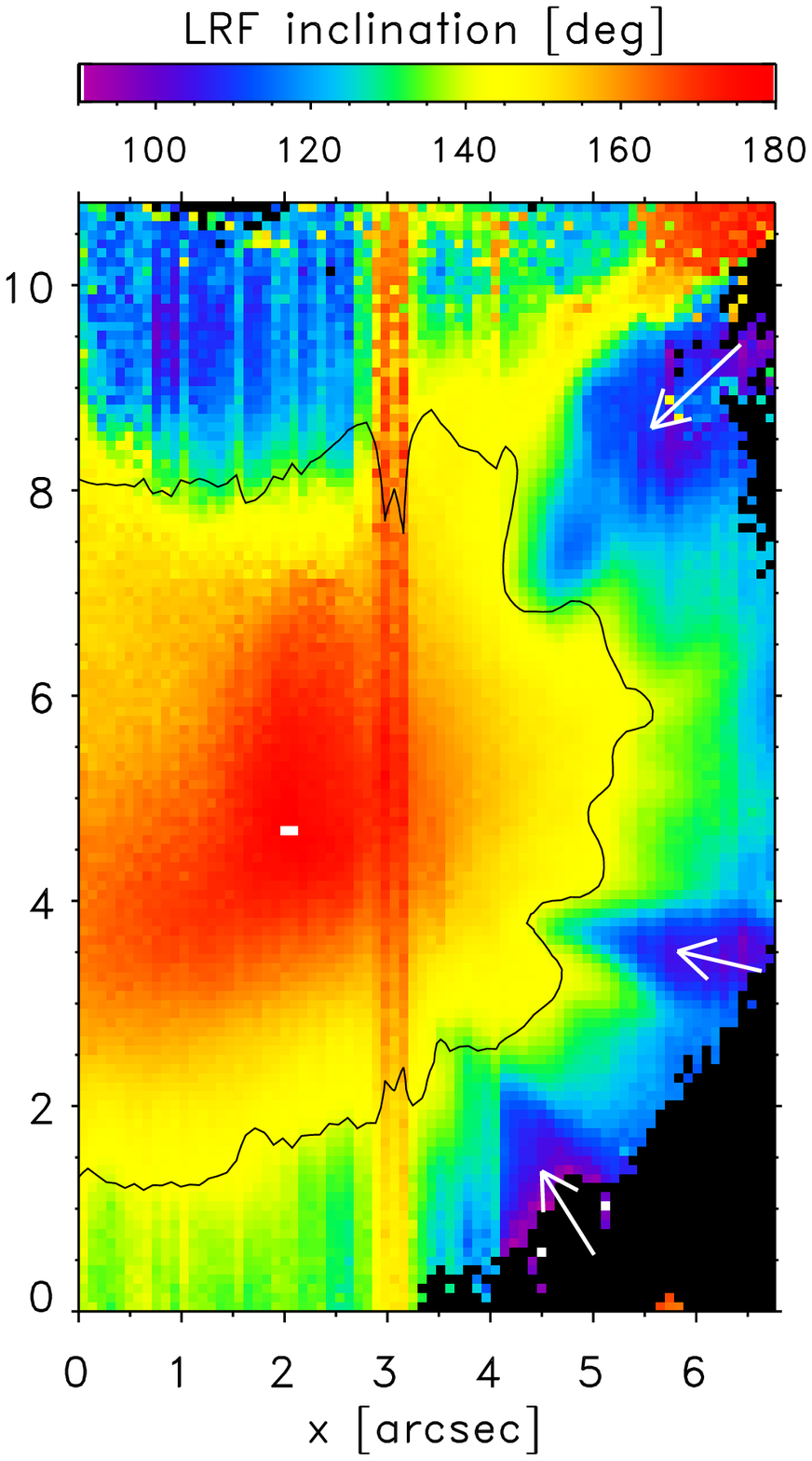}}
  \plotone{\figdir{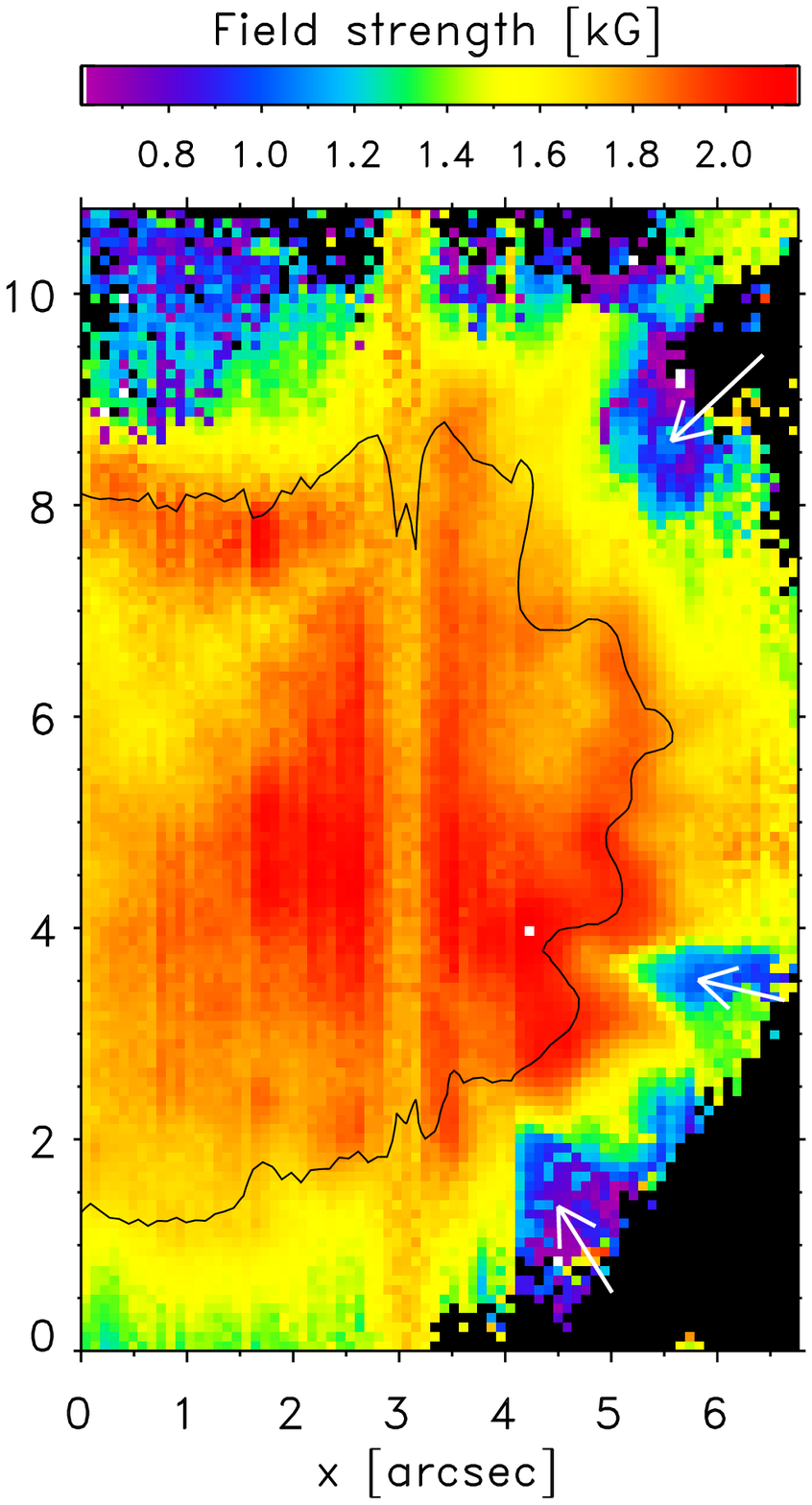}}
  \plotone{\figdir{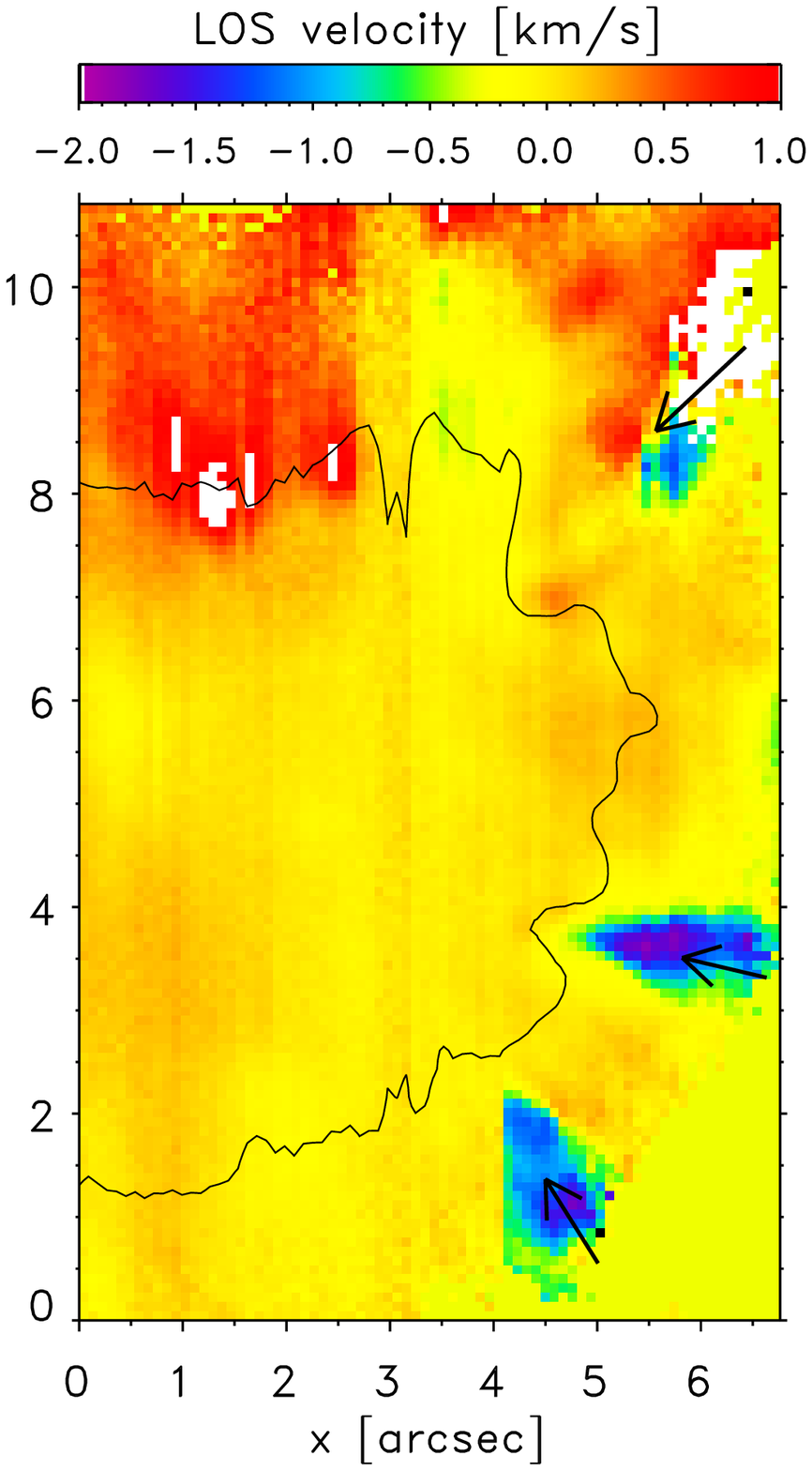}}
\caption{Magnetic structure of AR 10880 on May 10, 2006 ({\em top})
and May 11, 2006 ({\em bottom}). A different scale is used for each
day. From left to right: continuum intensity at 630.2~nm, field
inclination in the local reference frame, magnetic field strengh, and
LOS velocity maps. Black contours delimit the umbra.  Negative
velocities correspond to blueshifts. The position of weaker and more
inclined fields in the magnetic canopy is indicated with arrows.}
\label{fig:invmap} 
\epsscale{1.0} 
\vspace*{1em}
\end{figure*}

\section{Data analysis}\label{sec:analysis}
To get a qualitative idea of the dynamical and magnetic evolution of the
sunspot and its penumbra, we compute Dopplergrams and magnetograms from the
observed profiles as follows. LOS velocities are derived by applying a Fourier
phase method to the $I+V$ and $I-V$ signals of \FeI~630.25\,nm separately
\citep{schmidt+etal1999}. Magnetograms are created as the Stokes $V$ signal of
\FeI~630.25\,nm at $-9$~pm from line center. The magnetogram signal is roughly
proportional to $\cos \gamma$, with $\gamma$ the inclination of the vector
magnetic field to the LOS.

For a quantitative analysis of the observations we perform an
inversion of the Stokes profiles of the May 10 and May 11 data sets
using the SIR code \citep{ruizcobo+deltoroiniesta1992}.  In view of
the high angular resolution achieved by the instrument and
notwithstanding the variable conditions prevailing on May 11, we
regard a simple one-component atmosphere as sufficient to interpret
the spectra of the four lines observed with the polarimeter. The
inversion determines the three components of the vector magnetic field
(strength, inclination, and azimuth), the LOS velocity, and the macro-
and micro-turbulent velocities, all assumed to be height
independent. The initial temperature stratification is modified using
two free parameters.  This allows to change the temperature at optical
depth unity and the slope of the stratification. The code also
determines the amount of stray-light contamination. In total, nine
free parameters are retrieved from the inversion.  The vector magnetic
field returned by the code has been transformed to the local reference
frame (LRF), where inclinations are measured with respect to the local
vertical. The 180$^\circ$ ambiguity of the LOS field azimuth has been
solved by selecting the azimuth values that produce the more symmetric
magnetic field around the spot center in the LRF.

%
%

\section{Results}\label{sec:results}

Figure \ref{fig:obs} shows the sunspot decay on four consecutive days
corresponding to heliocentric angles of 13, 6, 15, and 32$^\circ$,
respectively.  The sequence starts on May 8 with the spot having a
fully developed penumbra (cf.\ the continuum intensity maps displayed
in the upper panels).  On May 9, the umbra splits in two parts and the
penumbra begins to disappear.  By May 10 the spot mainly consists of a
naked umbra, with some penumbral filaments NW of the main
umbra. Although the seeing conditions were very variable on May 11, no
trace of penumbral filaments can be detected in the continuum
intensity map. The only remnant of the spot on that day is a naked
umbra about 8\arcsec\/ wide. Therefore, the disappearance of the
penumbra took three days. Our May 13 observations show a small pore 
at the position of the spot which is no longer visible on May 14.

The middle panels of Fig.~\ref{fig:obs} display magnetograms of the spot. The
filamentary organization of the penumbra is conspicuous on May 8 and 9. On May
10, the penumbral structure has disappeared for the most part, and the
magnetic canopies of the naked umbrae are detected well beyond the edges of
the spot. On May 11, the magnetic canopy is very much reduced, showing three
distinct finger-like structures of positive polarity on the limb-side part of
the spot (the white features observed in the magnetogram to the SW and the
NW).  The fingers extend from the visible border of the umbra into the quiet
Sun by some 1-2\arcsec and have widths of about 0\farcs5.  To our knowledge,
such inhomogeneities have never been reported before, perhaps because of their
small sizes and/or short lifetimes.

The \CaII~K filtergrams in the bottom panels of Fig.~\ref{fig:obs}
show the sunspot penumbra as a dark structure at chromospheric
heights. The penumbra is rather homogeneous and do not exhibit
small-scale filaments.  Its size gradually decreases with time. 
Interestingly, however, on May 11 the spot still seems to have a
chromospheric penumbra which is better developed toward the S and NW.

Figure~\ref{fig:physmaps} displays a reconstructed G-band image of the spot on
May 11, together with the magnetogram showing the fingers, a Dopplergram, and
a \CaII~K filtergram.  As can be seen in the G-band image, the spot consists
only of a naked umbra without penumbral filaments. Very close to the umbral
border, on the west side of the spot, one can observe tiny G-band fibrils
cospatial with the fingers detected in the magnetogram. The fibrils show up as
bright elongated structures aligned radially, similar in shape and appearance
to the ones discovered by \citet{scharmer+etal2002}. The granulation around
the umbra is distorted and exhibits larger G-band intensities than the
surrounding quiet Sun. The Dopplergram reveals conspicuous blueshifts at the
position of the fingers, indicating plasma motions toward the observer. 
We want to emphasize that these blueshifts cannot be produced by Evershed 
flows, since on that part of the spot they would result in redshifted 
Stokes profiles.

The physical parameters retrieved from the inversion of the May 10 and
11 data sets are shown in Fig.~\ref{fig:invmap}.  We display continuum
images of the spot along with field inclination, field strength, and 
LOS velocity maps.  The arrows mark regions in the canopy where the
field is weaker and more horizontal. These regions look like fingers
and can be detected not only on May 11, but also on May 10 at position
angles where no penumbral filaments exist. Interestingly, the
penumbral filaments observed on the NW side of the spot on May 10 show
a typical pattern of spines and intra-spines \citep{lites+etal1993} in
the inclination map. This pattern is different from that associated
with the fingers.

Given the simplicity of our inversion, the atmospheric parameters
presented in Fig.~\ref{fig:invmap} should be regarded as averages
along the line forming region. The mean inclination of the canopy
fields at the border of the umbra is about 140$^\circ$ (corresponding
to 40$^\circ$ for a positive polarity spot), which agrees very well
with the values observed in pores \citep{keppens+martinezpillet1996}.
The finger-like inhomogeneities exhibit fields that are more
horizontal than the average by up to 20-40$^\circ$ at the same radial
distance. Also, the field in the fingers is weaker by up to
$\sim$600-1000~G. These differences are well above the uncertainties
in the retrieved parameters (some $5^\circ$ for the inclination and
100~G for the field strength). Figure~\ref{fig:invmap} demonstrates
that the fingers were observed on May 11 as opposite-polarity
structures only because of the large heliocentric angle of the
spot. Thus, the opposite polarities detected in the magnetograms do
not mean that the field is actually returning to the solar surface.

The kinematic configuration of the spot is displayed in the rightmost
panels of Fig.~\ref{fig:invmap}. On May 11, the LOS velocities in the
limb-side part of the canopy are nearly zero except for the fingers,
which show enhanced motions of 1-2~km~s$^{-1}$ toward the observer. 
These motions coincide with the regions of weaker and more horizontal 
fields. We do not find enhanced LOS velocities in the fingers of 
May 10, which might be due to either a less evolved state of the 
penumbra, or the smaller heliocentric angle of the spot.

\begin{figure*}
  \epsscale{.3}
  \plotone{\figdir{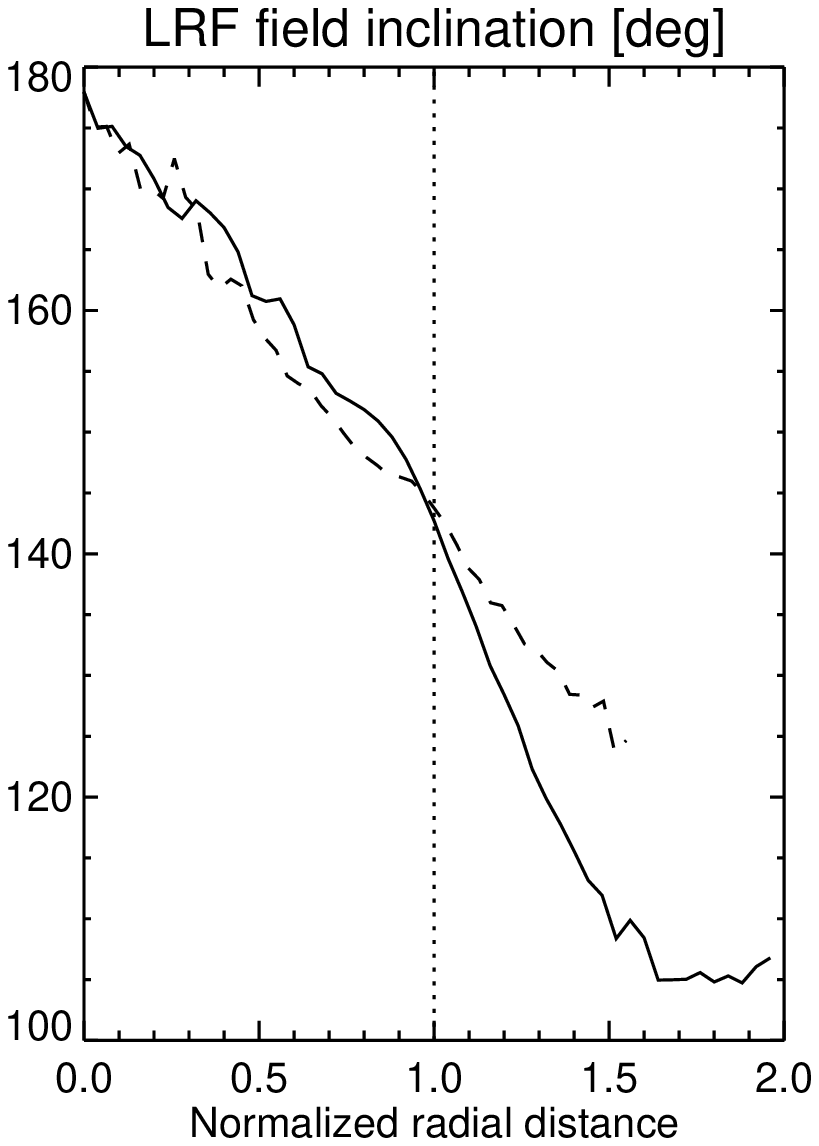}}
  \plotone{\figdir{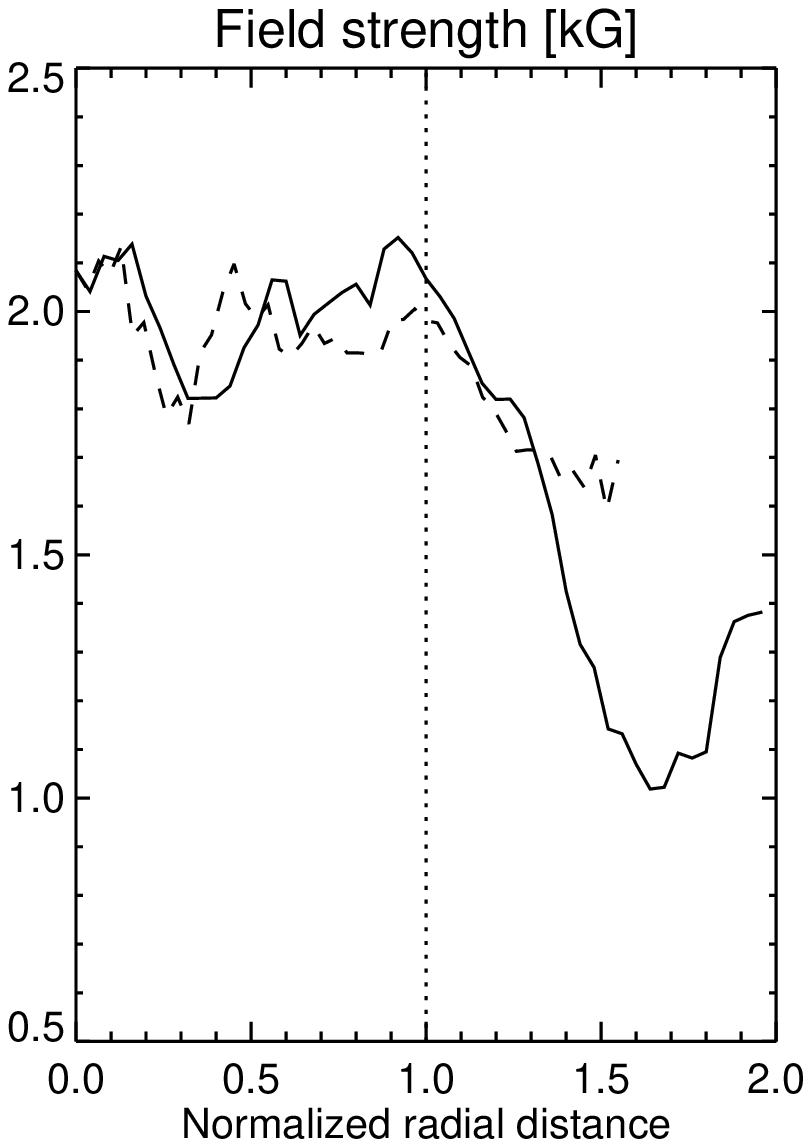}}
  \plotone{\figdir{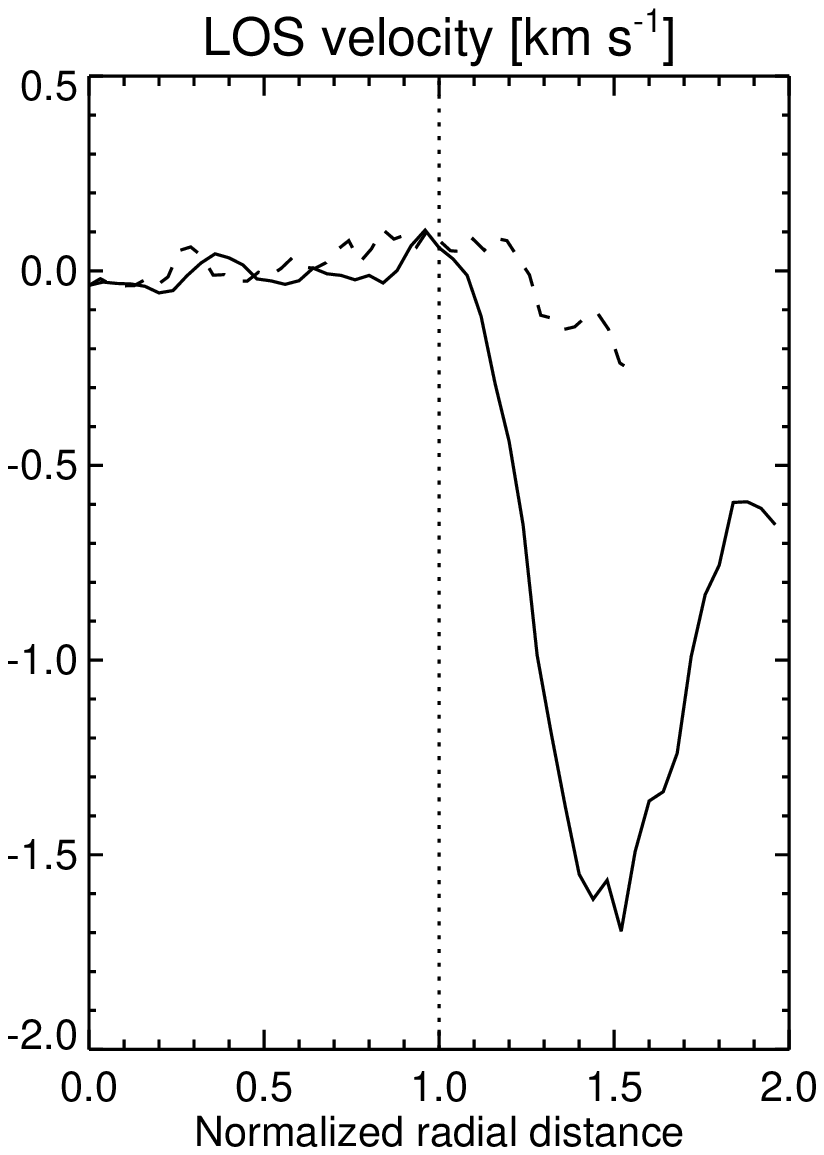}}
\caption{Radial variation of the field inclination in the LRF ({\em left}),
field strength ({\em center}), and LOS velocity ({\em right}) along the cuts
indicated in Fig.~\ref{fig:physmaps}. The solid lines represent cut A, which
samples one of the fingers, whereas the dashed lines give the atmospheric
parameters of the ``undisturbed'' magnetic canopy (cut B). The $x$-axis
indicate the normalized radial distance, with 0 the sunspot center and 1 the
border of the umbra.}
\label{fig:cuts}
\epsscale{1.0}
\vspace*{1em}
\end{figure*}

Figure~\ref{fig:cuts} illustrates the variation of the atmospheric
parameters along the two radial cuts marked in
Fig.~\ref{fig:physmaps}.  They sample one of the fingers (cut A) and
the nearby ``undisturbed'' magnetic canopy (cut B). As can be seen,
the field inclination decreases almost linearly from the center of the
umbra all the way to its visible border. From that point on, the
magnetic field becomes more inclined at a much faster pace in the
finger, while in the undisturbed canopy the inclination still
increases linearly. The finger reaches a maximum inclination of about
105$^\circ$ at radial distances of $r=1.5$ and beyond, while the
magnetic canopy has a field which is more vertical by up to 20$^\circ$
at the same position. In the umbra, the field strength and the LOS
velocity show values of around 2~kG and 0~km\,s$^{-1}$, respectively,
and do not change significantly with radial distance. Once the border
of the umbra is reached, however, the field strength decreases and the
LOS velocity increases dramatically in the finger. As a result, at $r
\sim 1.6$ the field strength goes down to $\sim$1 kG and the LOS
velocity reaches negative values of 1.7~km\,s$^{-1}$ (toward the
observer). In the magnetic canopy, the variation of field strength and
LOS velocity is much smaller.

\section{Discussion}\label{sec:discussion}
The physical properties of the small-scale inhomogeneities detected in the
magnetic canopy of the decaying sunspot are reminiscent of those of the
photospheric flux tubes that form the penumbra \citep{solanki2003,
borrero+etal2004, bellot2007}. Both have weaker and significantly more
horizontal fields than their surroundings. Unlike penumbral flux tubes,
however, the inhomogeneities are not associated with filamentary structures in
continuum intensity and do not harbor outward directed Evershed flows.

We interpret the scenario drawn by Fig.~\ref{fig:physmaps} as one 
of a magnetic canopy (i.e., horizontal fields) showing small-scale,
finger-like inhomogeneities of polarity opposite to that of the umbra.
The inhomogeneities are not associated with regular penumbral
filaments, suggesting that they are indeed located in a magnetic
canopy which is elevated above the continuum forming layers.  Below
the canopy, the plasma must be field-free for the most part, otherwise
dark structures would be observed in continuum intensity. As mentioned
before, the \CaII~K filtergram reveals a partly developed penumbra at
chromospheric heights. The existence of a penumbra in the chromosphere
is consistent with the scenario of an elevated magnetic canopy. The
fingers have no clear \CaII~K counterparts.

In principle, the blueshifts associated with the fingers could be
explained in a similar way as the classical Evershed effect. Already
\citet{meysch1968} proposed that the photospheric Evershed flow is the
result of a pressure difference between the two footpoints of
connected penumbral field lines: the inner penumbra displays field
strengths of around 1000~G, whereas typical photospheric magnetic
concentrations are found to harbor fields of around 1500~G. The
corresponding pressure difference between the two footpoints would
drive the flow \citep[cf.][and references therein]{thowei2004}. One
difficulty with this classical explanation of the Evershed effect 
is that we know now that the flow-carrying field lines return to the
photosphere not in typical magnetic flux concentrations, but right at
the penumbra-photosphere boundary (or even inside the penumbra) where
the field strength is indeed less than 1000~G \citep{west1997}. Yet,
an argument along those lines can be constructed here to explain the
observed blueshifts. As shown in Fig.~\ref{fig:invmap}, the field
strength near the decaying spot is about 2000~G, whereas in the outer
footpoint is only slightly above 1000~G. In this case, the siphon-flow
hypothesis favors an inward directed flow, in agreement with the
observations. We caution, however, that the gas pressure at the 
same heights in the two footpoints has not been infered in this work.

Another possibility is that the blueshifts represent upward motions of nearly
horizontal penumbral flux tubes. In the absence of strong Evershed flows, the
tubes might rise buoyantly to layers above the continuum forming region. This
would produce the disappearance of the penumbra in the photosphere, consistent
with the observation that the inhomogeneities occur in the last stages of the
spot evolution. The mechanism whereby the Evershed flow disappears in those
tubes remains unclear, but it might be related to the exhaustion of the mass
reservoir that feeds the flow.

The exact reason why all sunspots display an Evershed flow is not yet
known.  A gas pressure difference could be the origin of the flow
(either from a different field strengths at the two footpoints or from
an excess heat in the upflowing plasma, cf.~Schlichenmaier 2002), but
in any case considerable amounts of mass participate in this flow
during the typical lifetime of a sunspot. The existence of a mass
reservoir for the Eveshed flow must be understood in terms of the
exact connectivity of the sunspot field lines below the
photosphere. It has been suggested recently that sunspots are indeed
disconnected, at least in a dynamical sense, from their parent
toroidal flux tube \citep{schu2005}. Helioseismology also seems to
indicate that sunspots are a very shallow phenomenon \citep{zhao2001},
with flows in layers 10 Mm deep that cannot be related to an Evershed
flow source. Such a disconnection can also imply a disconnection of
the surface sunspot from the large mass reservoirs in the deep
interior, setting a maximum lifetime to the duration of the Evershed
flow for a given spot. 

\section{Summary}\label{sec:conclusions}
We have presented high angular resolution spectropolarimetric observations of
a decaying sunspot. The penumbra is seen to disappear from the photosphere
over the course of three days. When the spot looses its penumbra, magnetograms
in \FeI~630.25~nm show finger-like structures extending from the
visible border of the umbra into the quiet Sun. These inhomogeneities have
lengths of 1-2\arcsec and widths of about 0\farcs5. They are oriented radially
with respect to the spot center, but do not bear any relation with classical
penumbral filaments. The fingers exhibit conspicuous blueshifts, indicating
upward motions.

We have subject the observed Stokes spectra to simple one-component inversions
to infer the average physical properties of the fingers and their environs.
The umbra of the decaying spot is bigger in polarized light than in continuum
intensity, revealing the existence of a magnetic canopy. The fingers belong to
the canopy, but are found to possess weaker and more inclined fields. At the
same radial distance, the differences between the fingers and the undisturbed
magnetic canopy may amount to 600-1000 G and 40$^\circ$, respectively.  In
addition, the inversions confirm that the fingers harbor
line-of-sight velocities of up to 1-2 km\,s$^{-1}$.

The small sizes of these structures may explain why they have not been
described earlier. Their origin is unclear as yet, but they seem to be
associated with the disappearance of the penumbra. One possibility is that the
observed blueshifts are the signature of inward directed photospheric Evershed
flows, which would be driven by a classical siphon-flow mechanism.  Another
possibility is that the inhomogeneities represent penumbral field lines which no
longer carry outward directed Evershed flows. In the absence of the mass
associated with the normal Evershed flow, the field lines might rise to the
chromosphere by buoyancy, producing the disappearance of the penumbra at
photospheric levels. The upward motion of the field lines would be observed as
blueshifts. This scenario finds some support in the recent idea that sunspots
are disconnected from their parent toroidal flux tubes residing at the base of
the convection zone. If sunspots are indeed disconnected, they will be able to
store only a finite amount of mass. After some time (weeks to months) feeding
the Evershed flow, the mass reservoir of any sunspot would eventually be
exhausted and the flow would come to a halt.

To distinguish between scenarios, higher cadence observations and
longer coverage of decaying sunspots are required. This would make it
possible to characterize the time evolution of the vector field and to
determine the exact moment where the Evershed flow disappears.
Space-borne observations, virtually free from seeing effects,
represent the best option to solve this problem. In the future, we
plan to analyze spectropolarimetric measurements taken by Hinode
\citep{kosugi07} to shed more light on the nature of the fingers
described in this paper. The high resolution \CaII~H filtergrams of
Hinode will also allow us to investigate whether the disappearance of
the penumbra in the photosphere produces any effect at chromospheric
heights, as can be expected if the field lines rise buoyantly and
interact with the magnetic field of the chromosphere.

\acknowledgements The Diffraction-Limited Spectro-Polarimeter was built by NSO
in collaboration with the High Altitude Observatory of the National Center for
Atmospheric Research. Part of this work has been supported by the Spanish
Ministerio de Educaci\'on y Ciencia through projects ESP2006-13030-C06-02
and ESP2006-13030-C06-01.  This research has made use of NASA's
Astrophysics Data System (ADS).

%
%


\end{document}